\documentclass{article}
\usepackage{spconf,amsmath,graphicx}
\usepackage{CJKutf8}
\usepackage{algorithm}
\usepackage{algorithmicx}
\usepackage{algpseudocode}
\usepackage{amssymb}
\usepackage{hyperref}
\usepackage{cleveref}
\usepackage{makecell}
\usepackage{tabularx}
\usepackage{booktabs}
\usepackage{cite}

\title{EMALG: An Enhanced Mandarin Lombard Grid Corpus with Meaningful Sentences}
\name{Baifeng Li\footnotemark[2] \thanks{Thanks to XYZ agency for funding.}}
\address{Author Affiliation(s)}

\name{Baifeng Li$^1$${}^\dagger$, Qingmu Liu$^1$${}^\dagger$,Yuhong Yang$^{1,2*}$, Hongyang Chen$^1$, Weiping Tu$^1$, Song Lin$^3$}
\address{$^1$ National Engineering Research Center for Multimedia Software, School of Computer Science,\\
Wuhan University, Wuhan, China.\\
$^2$ Hubei Key Laboratory of Multimedia and Network Communication Engineering, \\
Wuhan, China.\\
$^3$ Guangdong OPPO Mobile Telecommunications Corp., China.}

\begin{document}
%
\maketitle
\renewcommand{\thefootnote}{}
\footnotetext{Our corpus is available at https://github.com/ASP-WHU/EMALG}
\renewcommand{\thefootnote}{\fnsymbol{footnote}}
\footnotetext[1]{Correspondence: yangyuhong@whu.edu.cn}

\begin{abstract}
This study investigates the Lombard effect, where individuals adapt their speech in noisy environments. We introduce an enhanced Mandarin Lombard grid (EMALG) corpus with meaningful sentences , enhancing the Mandarin Lombard grid (MALG) corpus. EMALG features 34 speakers and improves recording setups, addressing challenges faced by MALG with nonsense sentences. Our findings reveal that in Mandarin, meaningful sentences are more effective in enhancing the Lombard effect. Additionally, we uncover that female exhibit a more pronounced Lombard effect than male when uttering meaningful sentences. Moreover, our results reaffirm the consistency in the Lombard effect comparison between English and Mandarin found in previous research.
\end{abstract}

\section{Introduction}
\label{sec:intro}
\begin{figure}[htb]
    \centering
    \includegraphics[width=\linewidth]{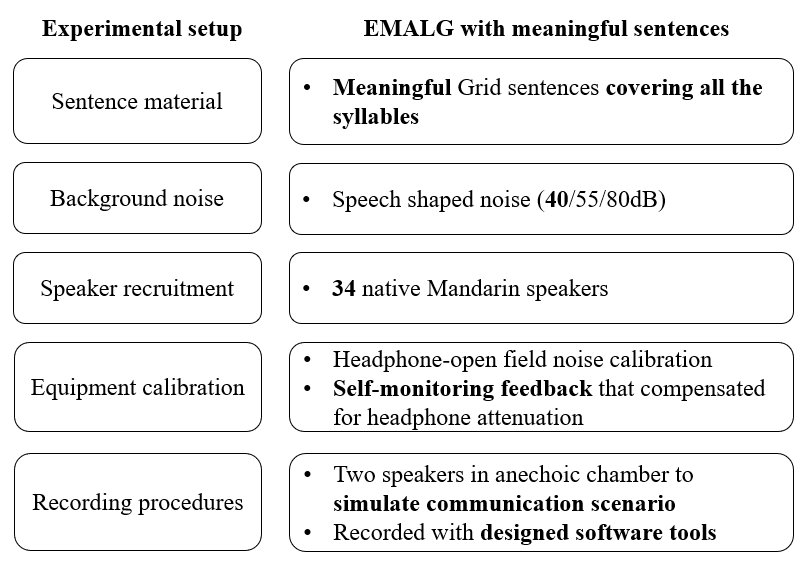} 
    \caption{Experimental setup of EMALG with meaningful sentences, the improvement is highlighted}
    \label{setup}
\end{figure}
The Lombard effect, denoting the unconscious adjustment of vocal patterns by individuals in noisy environments \cite{lombard1911signe}, plays a vital role in enhancing audibility and communication efficacy when faced with background noise.

Understanding the Lombard effect provides valuable insights into how our hearing and speaking adapt. Adaptive techniques can adjust speech automatically based on noise, making speech clearer in noise\cite{uma2021understanding}. This understanding can greatly improve systems like speech recognition, enhancement, and voice conversion\cite{skowronski2006applied,boril2009unsupervised,seshadri2019augmented,seshadri2019cycle}. 

Researchers have developed over 40 Lombard corpora in various languages, which includes English\cite{junqua1990acoustic,junqua1999lombard,cooke2006audio,huber2006effects,hansen2009analysis,cooke2010spectral,drugman2010glottal,vainio2012effect}, Polish\cite{kleczkowski2017lombard}, Hindi\cite{maheswari2020study}, Cantonese\cite{zhao2009effect}, Korean\cite{kim2005durational}, and Greek\cite{nicolaidis2012consonant}, tailored for Lombard effect exploration. These collections facilitate in-depth studies of the Lombard effect's acoustic and linguistic characteristics. Notably, Najwa Alghamdi et al. created an English Lombard Grid corpus\cite{alghamdi2018corpus}, featuring 54 speakers who recorded 5,400 utterances. This includes 2,700 Lombard speech at 80 dB SPL and 2,700 plain speech at 30 dB SPL. This dataset surpasses others in size, allowing for more robust handling of individual speaker effects.

Following the English Lombard Grid corpus, we introduced the Mandarin Lombard Grid (MALG) corpus\cite{yang2022mandarin}, with 28 speakers recording a total of 11,200 utterances. This includes 2,800 for each of three Lombard speech styles at 55, 70, 80 dB SPL and 2,800 plain reference utterances at 30 dB SPL. We examined the Lombard effect between the MALG and the English Lombard Grid corpus, revealing notable differences. These differences included larger increases in vowel-to-utterance ratio and F1 frequency and smaller increases in vowel duration and loudness. 

Despite its strengths, the MALG corpus has challenges, such as nonsense sentences and better data recording methods. Past research \cite{tasko2004variations} reveals that speaking nonsensical sentences causes notable variations in vocal movement speed, displacement, duration, and variability, potentially impacting the Lombard effect.

Recent research by Fei Chen et al. \cite{chen4330234understanding} revealed gender distinctions in the Mandarin Lombard effect, but it involved only 11 speakers. This small sample size makes it difficult to thoroughly characterize these effects across a diverse range of speakers.

This study aims to overcome these challenges through the creation of an enhanced MALG (EMALG) corpus. EMALG comprises 34 speakers (gender balanced), who recorded 10,200 meaningful utterances. Our goals are to bridge the gender gap in the Mandarin Lombard effect, investigate how nonsense sentences affect Lombard effect analysis, and reevaluate the Lombard effect comparison between English and Mandarin using EMALG.

\begin{table*}[th]
  \caption{Example sentences for meaningful (above) and nonsense (below) texts.}
  \label{sentence}
  \centering
  \begin{tabular}{cccccc}
    \toprule 
    \textbf{\makecell[c]{Name}} &
    \textbf{\makecell[c]{Verb}} &
    \textbf{\makecell[c]{Numeral}} &
    \textbf{\makecell[c]{Adjective}} &
    \textbf{\makecell[c]{Noun}} &
    \textbf{\makecell[c]{English translations}} \\
    \midrule 
    \begin{CJK}{UTF8}{gbsn}郑飞\end{CJK} & \begin{CJK}{UTF8}{gbsn}找到\end{CJK} & \begin{CJK}{UTF8}{gbsn}两个\end{CJK} & \begin{CJK}{UTF8}{gbsn}全新的\end{CJK} & \begin{CJK}{UTF8}{gbsn}饭盒\end{CJK}& Zhengfei found two brand-new lunch-box. \\
   Zhèng Fēi  & ZhǎoDào & LiǎngGè & QuánXīn de & FànHé \\
    \begin{CJK}{UTF8}{gbsn}青木\end{CJK} & \begin{CJK}{UTF8}{gbsn}种\end{CJK} & \begin{CJK}{UTF8}{gbsn}零个\end{CJK} & \begin{CJK}{UTF8}{gbsn}甜\end{CJK}& \begin{CJK}{UTF8}{gbsn}飞机\end{CJK}&Qingmu planted zero sweet airplane. \\
    Qīng Mù & zhòng & Líng Gè & Tián & FēiJī \\
    \bottomrule 
  \end{tabular}
\end{table*}

\section{Meaningful Mandarin Lombard Grid Corpus}
\label{sec:format}
EMALG has been specifically designed and collected in a two-speaker within an anechoic chamber to create a communication scenario to better induce the Lombard effect. In Figure \ref{setup}, we present the experimental setup of the EMALG dataset and highlighting the improvements of the new version. 
\subsection{Sentence Material}
\label{ssec:subhead}

Table \ref{sentence} compares meaningful and nonsense text examples in the study. The Grid sentence material is divided into categories such as names, verbs, quantifiers, adjectives, and nouns, each with ten variants. Missing phonemes from MALG\cite{yang2022mandarin} were filled in, and random word combinations were generated to create meaningful sentences. Each participant received a unique set of 100 sentences. The text design prioritized unique and evenly distributed word combinations to maximize the influence of phonemes on the Lombard effect.

\subsection{Background Noise}
\label{ssec:subhead}
Studies have indicated that the Lombard effect, when triggered by background noise bearing a spectral resemblance to speech noise, is more pronounced compared to its induction by background noise without speech frequency components\cite{alghamdi2018corpus}. Thus we chose the steady speech-shaped noise that better excites the Lombard effect.
Based on the findings from the Lombard effect categorization presented in \cite{yang2022mandarin}, no significant distinction exists between the intelligibility of speech produced in noise levels of 30dBA and 40dBA. Since 40dBA is more common than 30dBA in real scenarios, recordings were conducted at 40dBA, which we refer to as Plain style. The upper limit was set at 80dBA, considering it does not pose auditory harm when speakers were exposed briefly. Furthermore, a mid-range level of 55dBA (first Lombard speech style, L1), positioned between 40dBA (plain style) and 80dBA (second Lombard speech style, L2), was selected to facilitate a more detailed analysis of the Lombard effect's progression, referred to as first Lombard style (L1).
\begin{figure*}[htb]
  \begin{minipage}[b]{.24\linewidth}
    \centering
    \includegraphics[width=4.5cm]{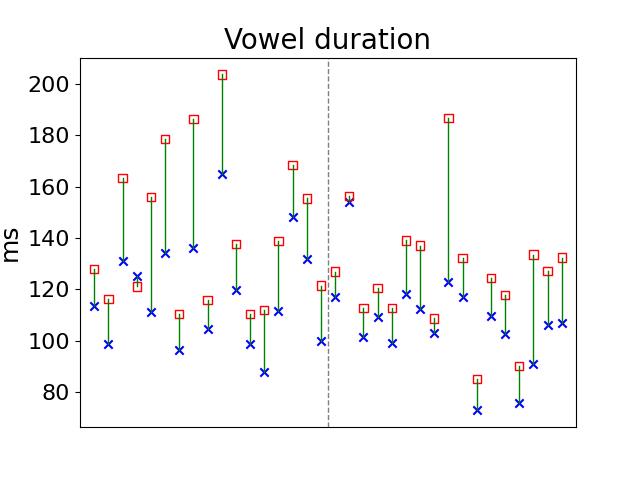}
  \end{minipage}%
  \hfill
  \begin{minipage}[b]{.24\linewidth}
    \centering
    \includegraphics[width=4.5cm]{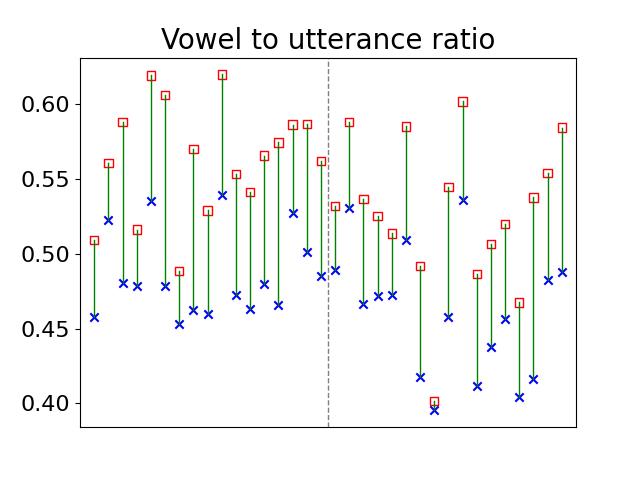}
  \end{minipage}%
  \hfill
  \begin{minipage}[b]{.24\linewidth}
    \centering
    \includegraphics[width=4.5cm]{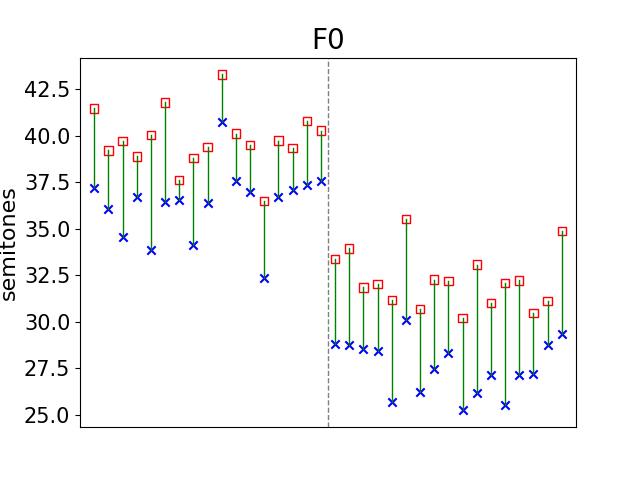}
  \end{minipage}%
  \hfill
  \begin{minipage}[b]{.24\linewidth}
    \centering
    \includegraphics[width=4.5cm]{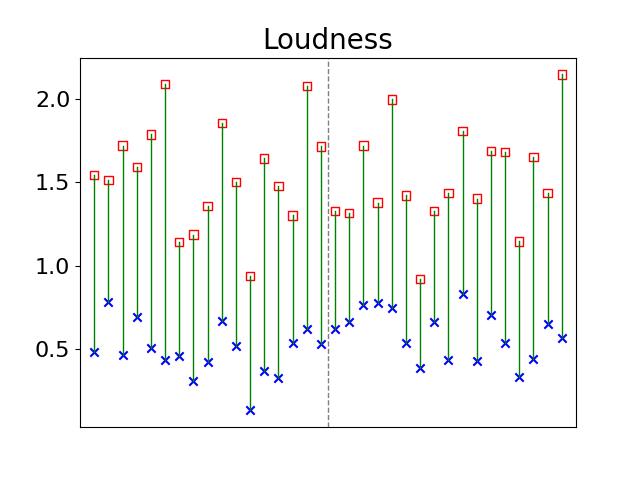}
  \end{minipage}%

  \begin{minipage}[b]{.24\linewidth}
    \centering
    \includegraphics[width=4.5cm]{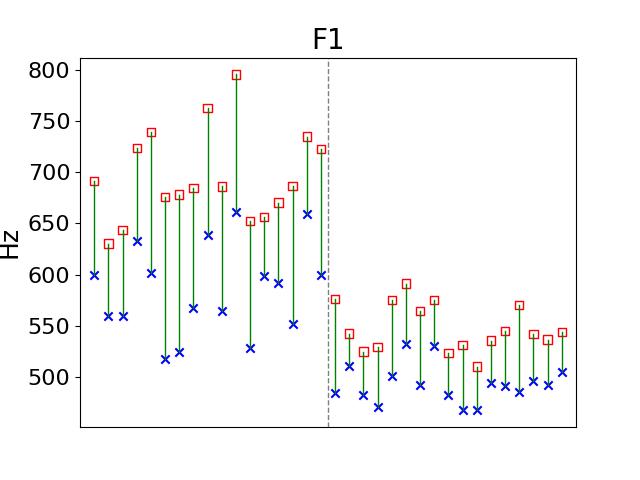}
  \end{minipage}%
  \hfill
  \begin{minipage}[b]{.24\linewidth}
    \centering
    \includegraphics[width=4.5cm]{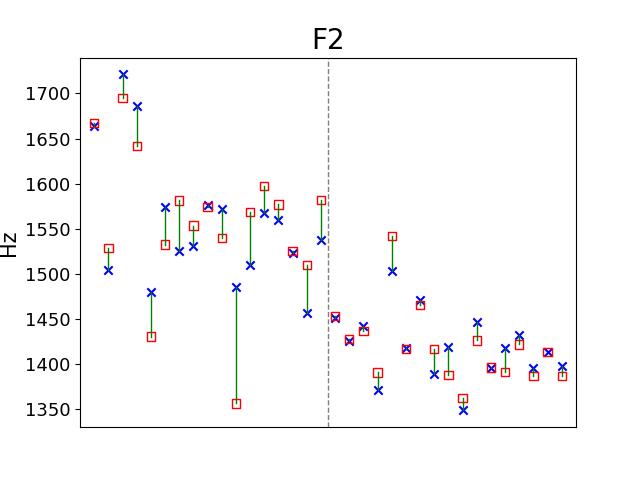}
  \end{minipage}%
  \hfill
  \begin{minipage}[b]{.24\linewidth}
    \centering
    \includegraphics[width=4.5cm]{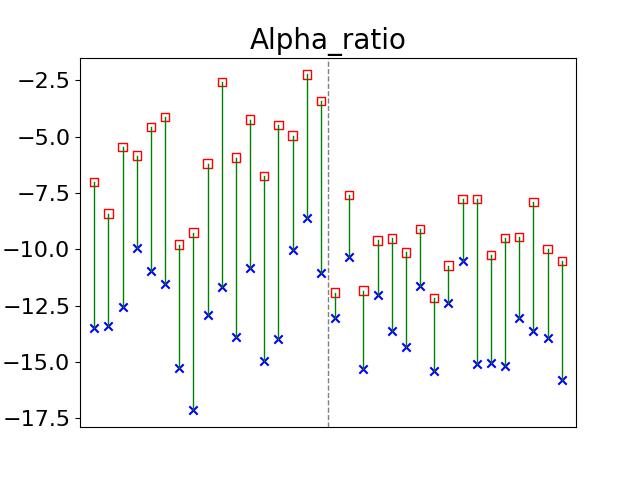}
  \end{minipage}%
  \hfill
  \begin{minipage}[b]{.24\linewidth}
    \centering
    \includegraphics[width=1cm]{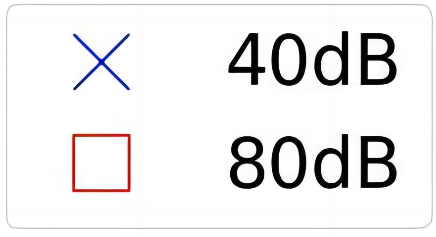}
  \end{minipage}%

\caption{Phonetic and acoustic parameters across talker: female speakers: left; male speakers: right}
\label{fig:res}
\end{figure*}

\begin{figure*}[htp]
  \begin{minipage}[t]{.25\linewidth}
    \centering
    \includegraphics[width=4.5cm]{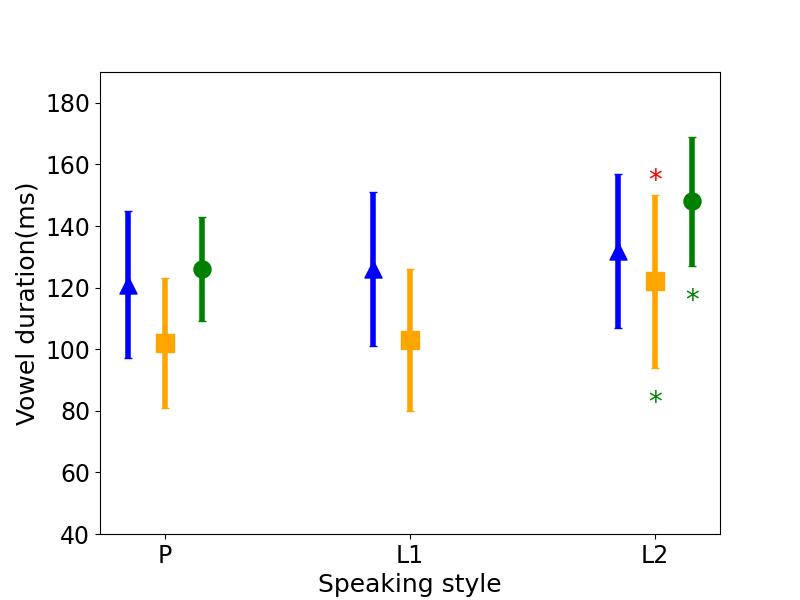}
  \end{minipage}%
  \hfill
  \begin{minipage}[b]{0.25\linewidth}
    \centering
    \includegraphics[width=4.5cm]{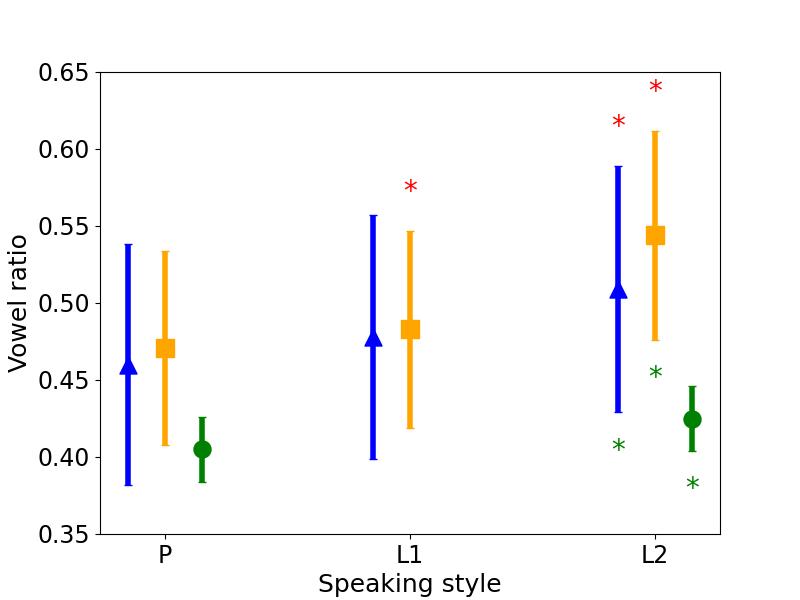}
  \end{minipage}%
  \hfill
  \begin{minipage}[b]{.25\linewidth}
    \centering
    \includegraphics[width=4.5cm]{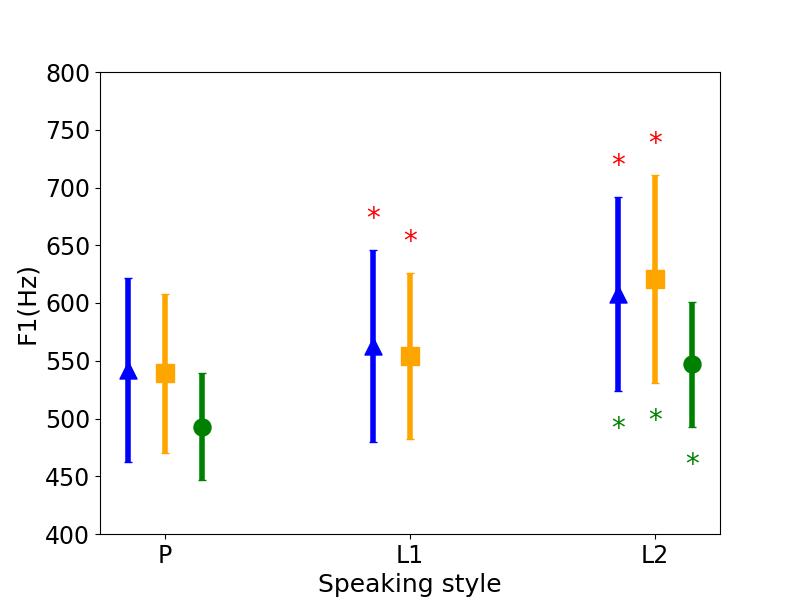}
  \end{minipage}%
  \hfill
  \begin{minipage}[b]{0.25\linewidth}
    \centering
    \includegraphics[width=4.5cm]{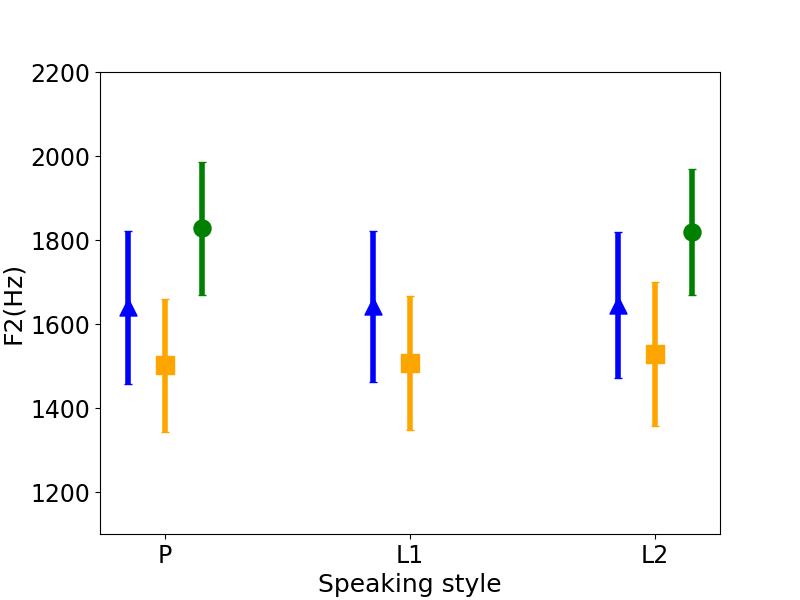}
  \end{minipage}%

  \begin{minipage}[t]{.24\linewidth}
    \centering
    \includegraphics[width=4.5cm]{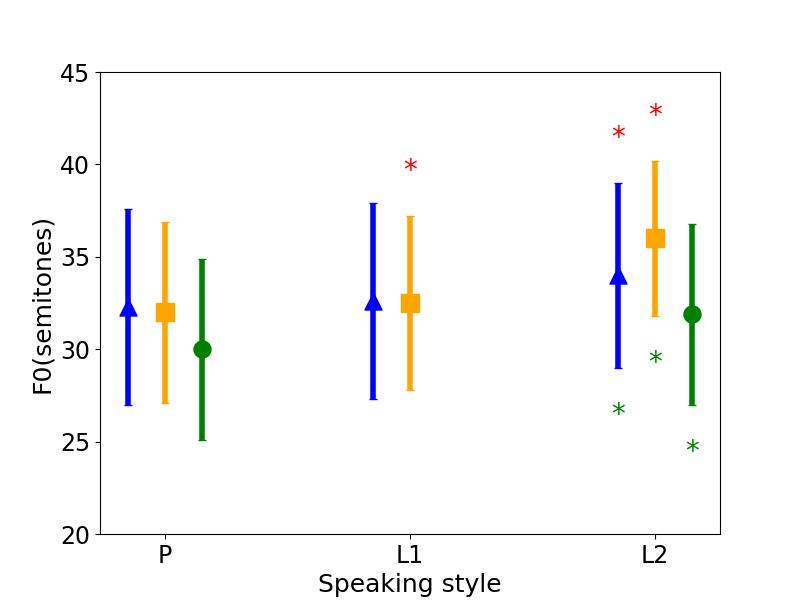}
  \end{minipage}%
  \hfill
  \begin{minipage}[t]{0.24\linewidth}
    \centering
    \includegraphics[width=4.5cm]{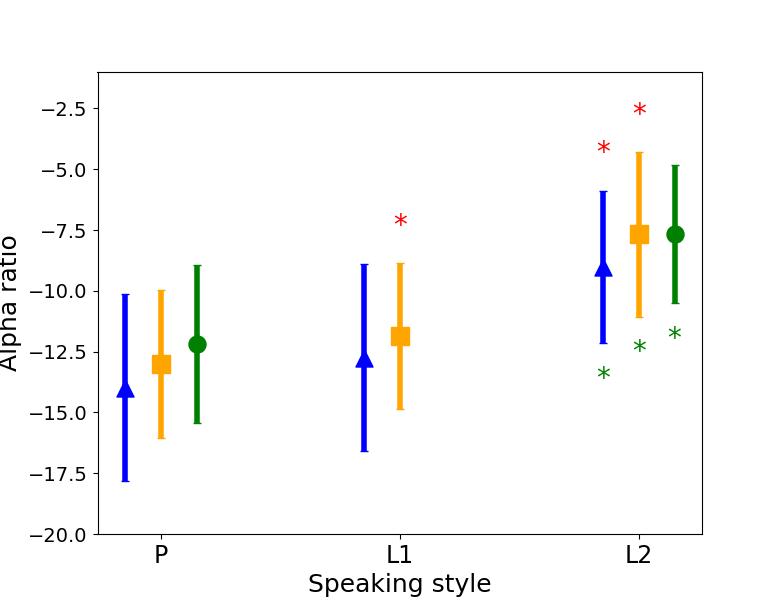}
  \end{minipage}%
  \hfill
  \begin{minipage}[t]{.25\linewidth}
    \centering
    \includegraphics[width=4.5cm]{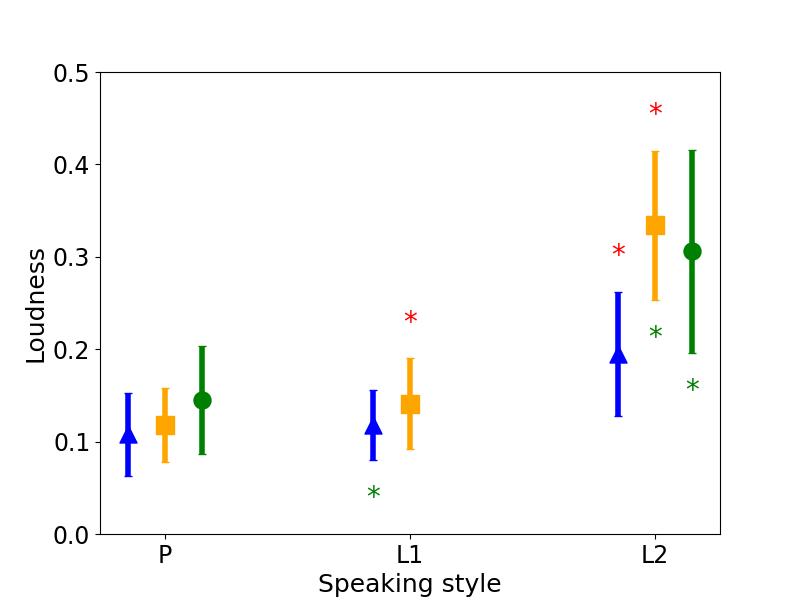}
  \end{minipage}%
  \hfill
  \begin{minipage}[t]{0.25\linewidth}
    \vspace{-2cm}
    \centering
    \includegraphics[width=4.5cm]{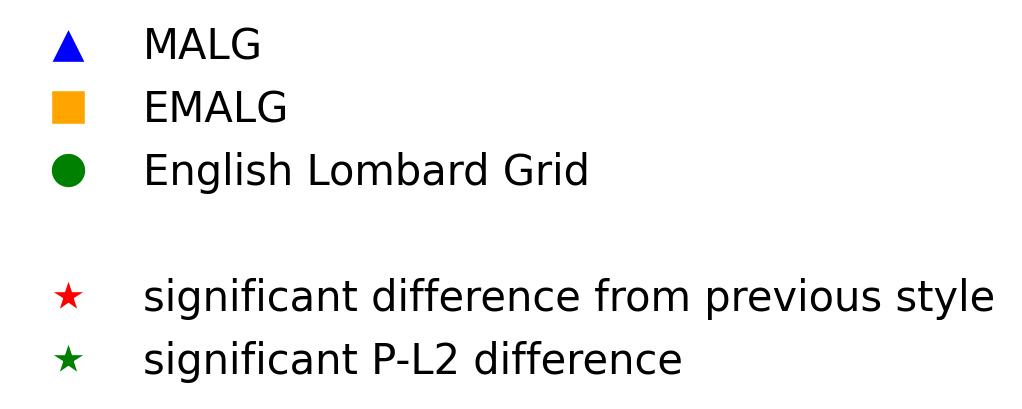}
  \end{minipage}%

\caption{Phonetic and acoustic modifications across three types of background noise levels (30/40 dBA, 55 dBA, 80 dBA) using the MALG\cite{yang2022mandarin}, EMALG corpora, and English Lombard Grid\cite{alghamdi2018corpus}. Speech styles are categorized as Plain (P) at 30 dBA for MALG and English Lombard Grid, and 40 dBA for EMALG; Lombard 1 (L1) at 55 dBA; and Lombard 2 (L2) at 80 dBA. For EMALG, loudness comparisons are standardized by reducing the audio energy by 20 dBA. Statistical significance was determined using a paired t-test, with the results showing p $<$ 0.001.}

\label{fig:parameter}
\end{figure*}

\subsection{Speaker Recruitment}
\label{ssec:subhead}
To ensure a comprehensive corpus from a sufficient number of speakers and account for individual variations in the Lombard effect, 34 speakers (17 male and 17 female) were recruited for this dataset collection. It was imperative that all participants were native Mandarin speakers and had passed the Level 2 or higher Mandarin proficiency test. Speakers are aged between 18 and 25 years. Each participant was tasked with reading 100 sentences in each of the 3 levels. All recordings took place in the anechoic chamber at Wuhan University. 

\subsection{Equipment Calibration}
\label{ssec:subhead}
We used the RODE NT1-A condenser microphone for signal acquisition. The audio signal was then routed through the ADA8200 digital-to-analog converter after being captured by the microphone and recorded on a computer equipped with the RME HDSPe RayDat sound card. Additionally, during the preparatory phase, we utilized the following hardware components: the G.R.A.S. 45BA KEMAR Head Torso, an LA-1440 sound level meter, a G.R.A.S. 42AA calibrator, Sennheiser HD300 closed-back headphones, and an array of twenty-one Bowers \& Wilkins speakers configured for surround sound.

The headphone and open-field noise calibration aimed to simulate open-field background noise using headphones. Initially, ambient noise was played through speakers to create a consistent sound field around the KEMAR Head \& Torso. This sound environment was recorded. Then, the background noise played through headphones while using the same Torso was also recorded. The energy difference between these two recordings was calculated to assess the disparity between the headphone-worn and open-field scenarios.

In real-world scenarios, human perceives sound through two primary pathways: bone conduction and air conduction. When wearing headphones, the air conduction pathway is obstructed. To compensate for headphone attenuation, we use TotalMix FX to channel a portion of the microphone's input to the headphones.  The experimenter wore headphones on one side and spoke to the microphone. By continually adjusting the gain from the microphone input to the headphones,  until there is no discernible difference between wearing and not wearing the headphones.

\section{Analysis of Lombard Effect}
\label{sec:pagestyle}

\subsection{Parameter Extraction}
\label{ssec:subhead}
To study the Lombard effect, we analyzed phonetic and acoustic parameters from plain and Lombard speech under three noise conditions. We computed the average duration of vowels as well as the ratio of total vowel duration to total vocalization duration. Formant frequencies F1 and F2 were estimated using Praat\cite{boersma2011praat}. Using openSMILE\cite{eyben2010opensmile}, we extracted key features from the Geneva Minimalistic Acoustic Parameter Set(GeMAPS)\cite{eyben2015geneva} set, including mean fundamental frequency (F0), loudness(an estimate of perceived signal intensity from an auditory spectrum), and the alpha ratio (energy ratio between 50-1000 Hz and 1-15 kHz).

\subsection{Lombard Effect Analysis Of The Mandarin Corpus Across Talker}
\label{ssec:subhead}
Figure \ref{fig:res} illustrates the mean values of seven parameters for various speakers at plain and Lombard style. While most of the speakers exhibit consistent trends of variation, the magnitudes of these changes are not uniform across all speakers.

Evident gender differences exist in F0, F1, and Alpha ratio. In the transition from Plain style to Lombard style (L2), male exhibit a 35\% greater average increase in F0 compared to female (4.6 for male, 3.4 for female). However, physiological factors lead to a lower fundamental frequency in male under the Plain style, potentially amplifying this increase. Additionally, male show an average increase of 3.8 in Alpha ratio, while female experience an 81\% greater increase (6.9 for female, 3.8 for male). In terms of F1, female demonstrate a remarkable 101\% greater average increase than male (110.3 for female, 54.8 for male), emphasizing a stronger Lombard effect in female.

\subsection{Lombard Effect Comparison Between MALG And EMALG}
\label{ssec:subhead}

Figure \ref{fig:parameter} reveals the differences of the Lombard effect between nonsense and meaningful Mandarin corpora in terms of phonetic and acoustic parameters, the statistical comparison are as follows:
(1) Between P and L1, significant parameter variations are observed in EMALG for vowel ratio, F1, F0, loudness, and alpha ratio, while in MALG, only F1 shows significant changes.
(2) Between L1 and L2, significant parameter variations are found in EMALG for vowel duration, vowel ratio, F1, F0, loudness, and alpha ratio, while in MALG, significant changes are observed for vowel ratio, F1, F0, loudness, and alpha ratio.
(3) Between P and L2, EMALG and MALG exhibit consistency, with all parameters except F2 showing significant variations.
These variations suggest that the Lombard effect is stronger in meaningful sentences.

\subsection{Lombard Effect Comparison Between The Mandarin And English Corpus}
\label{ssec:subhead}
To examine the Lombard effect across diverse languages, a comparative analysis was undertaken among the Phonetic and Acoustic parameters of MALG, EMALG and the English Lombard Grid. The results of EMALG align with MALG, except for F2. In contrast to English, the duration of vowels in Mandarin remains generally consistent with English. However, Mandarin exhibits a higher proportion of vowels and is more influenced by the Lombard effect, showing a greater increase compared to English. This observation may be attributed to Mandarin being a tonal language, where different tones on vowels convey distinct semantic meanings. Moreover, in Mandarin, consonant duration tend to be shortened during periods of stronger Lombard effect.

F1 frequency in Mandarin escalated more substantially, likely due to speakers’ tendency for wider lip opening during strong vocal effect. Conversely, a higher F2 frequency in English indicates a forward tongue position.

\section{Conclustion}
\label{sec:typestyle}
This paper presents a Mandarin Lombard Grid corpus, which includes 34 speakers and 10,200 sentences covering plain and Lombard speech at different sound levels (40dBA, 55dBA, and 80dBA). The study investigates the Lombard effect in both meaningful and nonsense sentences. The findings reveal a stronger Lombard effect in meaningful text, characterized by increased vowel duration, reduced vowel ratio, and a higher alpha ratio, indicating a rise in high-frequency energy relative to low-frequency components in speech. Additionally, our study indicates that when using meaningful sentences, female speakers not only exhibited an increase in the alpha ratio\cite{yang2022mandarin} but also showed growth in F1.

Compared to the English Lombard Grid corpus, the results generally align. There are significant differences in all parameters except for F2. The divergence of vowel pronunciations reveals distinct phonetic adaptations of Mandarin and English speakers in response to noise, with Mandarin emphasizing vowel clarity due to its tonal nature, and English focusing on stress patterns and intonation.

The Lombard effect study aims to leverage human vocal traits for enhancing speech intelligibility in real-world applications. EMALG has the potential to improve enhancing speech recognition, improvement, and separation systems. Upcoming research includes analyzing how various noise types affect the Lombard effect in the current dataset and comparing the Lombard effect in the meaningful Grid corpus with the natural TIMIT corpus.

\section{Acknowledgments}
This research is funded in part by the National Natural Science Foundation of China (62171326, 62071342), Key Research and Development Program of Hubei Province (220171406) and Guangdong OPPO Mobile Telecommunications Corp.

\small
\bibliographystyle{IEEEbib}
\bibliography{refs.bib}

\end{document}